\documentclass[aps,prl,reprint,superscriptaddress]{revtex4-1}
\pdfoutput=1
\usepackage{graphicx}
\usepackage{dcolumn}
\usepackage{hyperref}
\usepackage{bm}
\usepackage{epstopdf} 
\usepackage{color}
\usepackage{SIunits}
\usepackage{ulem}

\newcommand{\bra}[1]{\langle #1|}
\newcommand{\ket}[1]{|#1\rangle}

\newcommand{\bb}[1]{\left( #1 \right)}

\newcommand{\Sz}{\hat{S}_z}

\newcommand{\Kone}{K_1}
\newcommand{\KtwoAB}{K^{(2)}_{12}}
\newcommand{\KtwoBB}{K^{(2)}_{22}}

\newcommand{\cgpe}{2C-GPE}
\newcommand{\V}{\mathcal{V}}

\begin{document}
\preprint{APS/123-QED}
\title{Fundamental Limit of Phase Coherence in Two-component Bose-Einstein Condensates} 
\author{Yifan Li}
\affiliation{Department of Physics, University of Basel, Switzerland}
\affiliation{Department of Engineering Physics, Tsinghua University, China}
\author{Krzysztof~Paw{\l}owski}
\affiliation{Center for Theoretical Physics, Polish Academy of Sciences, Warsaw, Poland}
\author{Boris D\'ecamps}
\author{Paolo Colciaghi}
\author{Matteo Fadel}
\author{Philipp Treutlein}
\author{Tilman Zibold}
\email[]{tilman.zibold@unibas.ch}
\affiliation{Department of Physics, University of Basel, Switzerland}
\date{\today}

\begin{abstract}
We experimentally and theoretically study phase coherence in two-component Bose-Einstein condensates of $^{87}{\rm Rb}$ atoms on an atom chip. Using Ramsey interferometry we determine the temporal decay of coherence between the $ |F=1, m_{F}=-1\rangle$ and $|F=2, m_{F}=+1\rangle$ hyperfine ground states. We observe that the coherence is limited by random collisional phase shifts due to the stochastic nature of atom loss. The mechanism is confirmed quantitatively by a quantum trajectory method based on a master equation which takes into account collisional interactions, atom number fluctuations, and losses in the system. This decoherence process can be slowed down by reducing the density of the condensate. Our findings are relevant for experiments on quantum metrology and many-particle entanglement with Bose-Einstein condensates and the development of chip-based atomic clocks.
\end{abstract}

\maketitle
Atomic Bose-Einstein condensates (BECs) in magnetic traps are exceptionally well-isolated quantum many-body systems. Distinct from most other systems, the coherence of atomic BECs in state-of-the-art experiments is limited by intrinsic dynamics rather than interactions with the environment~\cite{Yvan2013,Yvan2018}. Elucidating the dominant decoherence mechanisms is thus an intriguing challenge for quantum many-body physics, but at the same time also highly relevant for applications of BECs in quantum metrology~\cite{pezze2018} and in experiments on the foundations of quantum physics~\cite{2016Bell, Matteo2018, 2018Oberthaler, Klempt2018}. 

The spatial coherence across a condensate has been studied in different systems~\cite{1999BEC, bloch2000, 1D2007, 1D2011}. Here we focus on the temporal coherence of three-dimensional trapped BECs, which limits the preparation fidelity of non-classical states and the duration of interferometric sequences in precision measurements. Theoretical studies of BEC temporal coherence have suggested different decoherence mechanisms resulting from the interplay of elastic collisional interactions, inelastic collisional atom loss, and finite temperature~\cite{Yvan2013,Yvan2018}. Previous experiments have mostly investigated inhomogeneous dephasing due to mean-field driven spatial dynamics of the condensate wavefunctions ~\cite{hall1998, egorov2011, 2018phase}, while homogenous dephasing was attributed to technical noise. Inhomogeneous dephasing and spatial demixing can be reversed by spin echo techniques ~\cite{2018phase}. Other experiments investigated the effect of phase noise on spin-squeezed states \cite{laudat2018}. However, a detailed experimental study of the predicted homogeneous phase decoherence mechanisms \cite{Yvan2013,Yvan2018}, which fundamentally limit the BEC coherence, has not yet been reported.

\begin{figure}[hb]
\includegraphics[width=8.6cm]{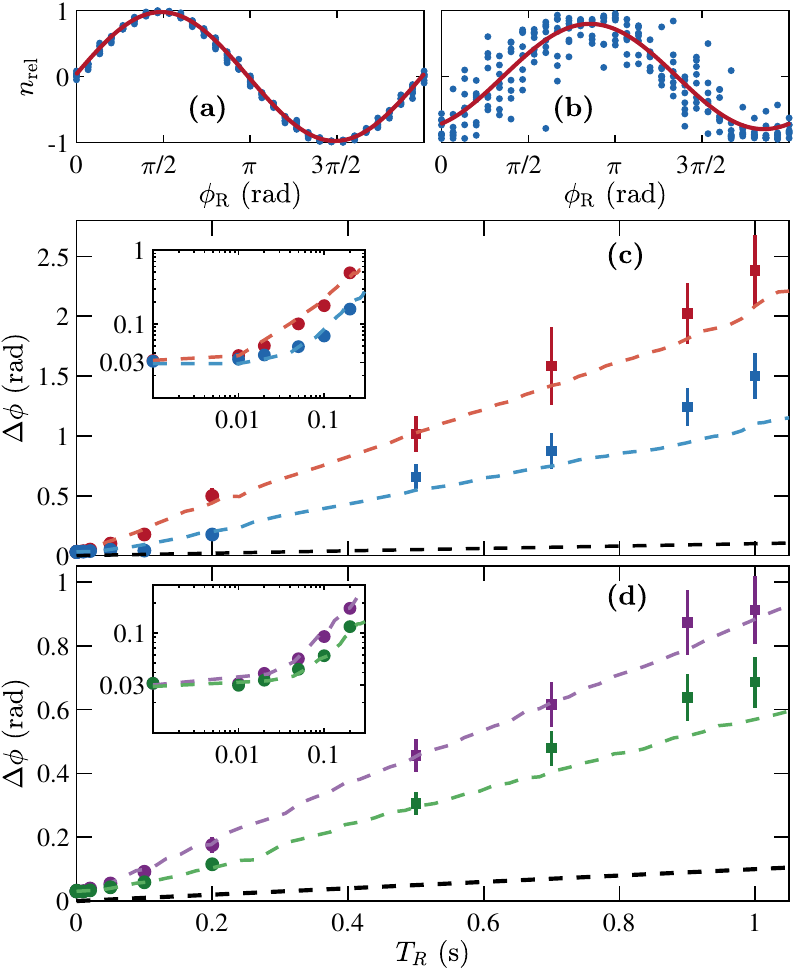}
\caption{\label{fig:main-result}
Ramsey interferometer fringes at $T_{R}=\unit{0.1}{ms}$ (a) and $T_{R}=\unit{200}{ms}$ (b), with red curve showing a sinusoidal fit of the experimental data (blue dots). The fringe at \unit{0.1}{ms} shows almost no sign of phase noise while after \unit{200}{ms} the fringe is smeared out by phase noise and the contrast is reduced. Phase noise as a function of $T_{R}$ is shown for a tight trap with $\omega_{(r,z)}=2\pi\times(714,114) \unit{}{Hz}$ (c) and a shallow trap with $\omega_{(r,z)}=2\pi\times(301,113) \unit{}{Hz}$ (d). Experimental results are shown before (red and purple) and after collisional phase shift correction (blue and green). Dots and square symbols represent data at $T_{R}<\unit{0.2}{s}$ and $T_{R}>\unit{0.2}{s}$ respectively, evaluated with different methods (see text). Results of quantum trajectory simulations are given by the dashed line with corresponding color (statistical uncertainty too small to be shown). Black dashed line shows the upper bound on technical noise (see text). The insets show the phase noise at $T_{R}<\unit{0.2}{s}$ in a log-log scale. The minimal phase uncertainty is close to the projection noise of a coherent spin state $\Delta \phi=1/\sqrt{N}\approx\unit{0.033}{rad}$.}
\end{figure}

We report experiments and corresponding theoretical simulations of the decoherence mechanisms in two-component BECs of $^{\rm 87}$Rb atoms. Using Ramsey interferometry, we explore the phase coherence of the hyperfine ground states $|1\rangle\equiv|F=1,m_{F}=-1\rangle$ and $|2\rangle\equiv|F=2,m_{F}=+1\rangle$ in a magnetic trap on an atom chip~\cite{bohi2009,Riedel2010,Caspar2013}. This system is used in recent experiments on many-particle entanglement and quantum metrology \cite{pezze2018} as well as compact atomic clocks~\cite{TACC, laudat2018}. Different aspects of coherence in this system have been studied, both in thermal and degenerate ensembles~\cite{hall1998, sortais2000, harber2002, treutlein2004, Rosenbusch2010, Klempt2011, egorov2011, Rosenbusch2012, TACC, laudat2018, 2018phase}. In our setup, high atom number stability and detection resolution allow us to suppress the effect of preparation imperfections to the point where intrinsic decoherence mechanisms dominate. We observe a temporal increase of relative phase fluctuations between the two hyperfine states (Fig.~\ref{fig:main-result}), which is primarily caused by random collisional phase shifts due to the stochastic nature of atom loss. This explanation is confirmed quantitatively by a theoretical model based on the quantum trajectory approach, which takes into account elastic collisional interactions, collisional atom loss, initial atom number fluctuations, and several other technical noise sources. We conclude that the phase coherence of two component BECs, which are inherently plagued by atom loss due to their chemical metastability, is fundamentally limited by this decoherence mechanism \cite{Yvan2018}.

In our experiment, a BEC of $1020\pm40$ $^{87}$Rb atoms is prepared in the hyperfine state $|1\rangle$ and trapped in a cigar-shaped magnetic potential with harmonic trapping frequencies $\omega_{(r,z)}=2\pi\times\unit{(714,\ 114)}{Hz}$. We coherently couple the two hyperfine states $|1\rangle$ and $|2\rangle$ via a resonant two-photon transition induced by microwave and radio-frequency magnetic fields, with a two-photon Rabi frequency of $\approx2\pi\times\unit{600}{Hz}$ at a detuning of $\approx2\pi\times\unit{500}{kHz}$ from the intermediate state $|F=2,m_{F}=0\rangle$. The two states $|1\rangle$ and $|2\rangle$ are of special interest because their first order differential Zeeman shift vanishes at a magnetic field of $\approx\unit{3.23}{G}$, making superpositions of the two states largely insensitive to magnetic field fluctuations~\cite{harber2002}. In principle, spin-dependent interactions can lead to demixing of the two components~\cite{hall2007}. However, this effect is weak in our system because of low atom density, as confirmed by a two-component Gross-Pitaevskii equation (2C-GPE) simulation and independent experiments~\cite{Riedel2010,Caspar2013}. The atoms can thus be treated as condensed in a single spatial mode and can be regarded as pseudo-spin 1/2. The whole ensemble can then be well described in terms of a collective spin~\cite{pezze2018}.

In order to investigate phase coherence between the two spin states, we perform Ramsey interferometry. Starting with a BEC in state $|1\rangle$, we apply a $\pi/2$ pulse to prepare an equal superposition coherent spin state. After an interrogation time $T_{R}$, a second $\pi/2$ converts the accumulated phase into an observable atom number difference. We measure the atom numbers in the two spin states $N_{1}$ and $N_{2}$ by absorption imaging, and evaluate the normalized atom number imbalance $n_{\text{rel}}=(N_{1}-N_{2})/(N_{1}+N_{2})$. By scanning the relative phase $\phi_{R}$ of the two pulses we record the Ramsey interferometer fringe in phase domain (Fig.~\ref{fig:main-result}, top panel). The outcome of such an experiment can be described by the following heuristic model
\begin{equation}
n_{\rm rel}(\phi_{R})=\V\sin(\phi_{R}+\phi),
\label{eq:nrel}
\end{equation}
where $\V$ is the fringe visibility, and $\phi$ is a classical random variable describing the phase accumulated by the state during the interrogation time $T_{R}$, which fluctuates due to noise. The phase noise quantified by the standard deviation $\Delta \phi$ is evaluated in the following way.

For short Ramsey times $T_{R}<\unit{0.2}{s}$, we experimentally set the Ramsey phase to the value $\phi_{R}=\phi_{0}$ where the mean of $n_{\rm rel}$ is zero. At this point, with the maximal slope of the fringe $\partial n_{\rm rel}/\partial\phi=\V$, the phase is evaluated as $\phi=\arcsin(n_\text{rel}/\V)$. In general, $\V<1$ due to the asymmetric losses in the two spin states. We repeat the experiment to gather statistics and evaluate the standard deviation $\Delta\phi$ assuming a Gaussian distribution of the phase fluctuations.
 
For long Ramsey times $T_{R}>\unit{0.2}{s}$, the above method is not reliable due to ambiguities of the phase. As a more reliable evaluation, we fit the entire Ramsey fringe with $n_{\text{rel}}(\phi_{\text{R}})=C\sin(\phi_{\text{R}}-\phi_{0})$, where $C$ is the contrast, and $\phi_{0}$ is the fitted phase at zero-crossing. For a Gaussian distribution of the phase fluctuations, $\Delta \phi$ is evaluated by the relation $C=\V\exp({-\frac{1}{2}\Delta\phi^{2}})$. 

We observe an increase of the phase noise with time (Fig.~\ref{fig:main-result}). The main interest of this work is to investigate the origins of these fluctuations, which can be divided into technical and intrinsic sources of noise. On the technical side, the system is suffering from detection noise, local oscillator noise, and magnetic field fluctuations. Low detection noise in our experiment, mostly due to photon shot noise in the absorption images, allows atom number detection below the spin projection noise ~\cite{Riedel2010}. The standard deviation in $N_1$ and $N_2$ is $\sigma_{\text{det,1}} \approx\ 3.6$ and $\sigma_{\text{det,2}}\approx 4.1$ for the two states, respectively, resulting in an equivalent phase noise of $\Delta\phi_\text{det} \approx 5.3\times10^{-3} \unit{}{rad}$ for $T_R=0$ and $\Delta\phi_\text{det} \approx 1.9\times10^{-2} \unit{}{rad}$ for $T_R=\unit{1}{s}$, negligible compared to other noises. Local oscillator noise and fluctuations of the magnetic field will directly effect the phase between the two Ramsey pulses. To estimate the contribution of these technical noises, we perform the same experiments with non-condensed atoms in a relaxed trap with trap frequencies $(\omega_{r},\omega_{z})=2\pi\times(301,113)\unit{}{Hz}$. The non-condensed atoms suffer from the same technical noise, but due to a significantly reduced atomic density, the effects of interactions are largely suppressed. The results of this experiment can be well fitted with a linear model $\Delta\phi_{\text{tech}} (t)=\unit{0.1}{(rad/s)}\times t$ for $T_R<\unit{12}{s}$. The fit is used as an upper bound on the technical noise in our experiment, shown in Fig.~\ref{fig:main-result} with a black dashed line. This demonstrates that technical noise contributes little to the observed phase noise in our experiments. 

Intrinsic to the two-component BEC, the collisional interactions also introduce phase fluctuations. The dynamics of the collective spin due to elastic collisions between atoms can be described by the Hamiltonian \cite{li2009}:
\begin{equation}
\label{eq:ham}
\hat{H}= \hbar\tilde{\chi}\hat{N}\Sz+\hbar\chi\Sz^2,
\end{equation}
where $\hat{S}_z =({\hat{N}_1-\hat{N}_2})/2$ is the $z$ component of the collective spin operator, and $\hat{N}=\hat{N_1}+\hat{N_2}$ is the total atom number. The parameters $\tilde{\chi}=\frac{1}{2\hbar}\bb{\frac{\partial^2 E}{\partial N_{1}^2}-\frac{\partial^2 E}{\partial N_{2}^2} }$ and $\chi=\frac{1}{2\hbar}\bb{\frac{\partial^2 E}{\partial N_{1}^2}+\frac{\partial^2 E}{\partial N_{2}^2}-2\frac{\partial^2 E}{\partial N_{2}\partial N_{1}}}$ depend on the energy $E$ of the two interacting components with mean atom number $\bar{N}_1$ and $\bar{N}_2$ in state $|1\rangle$ and $|2\rangle$ respectively \cite{li2009}. 

\begin{figure}[b]
\includegraphics[width=8.6cm]{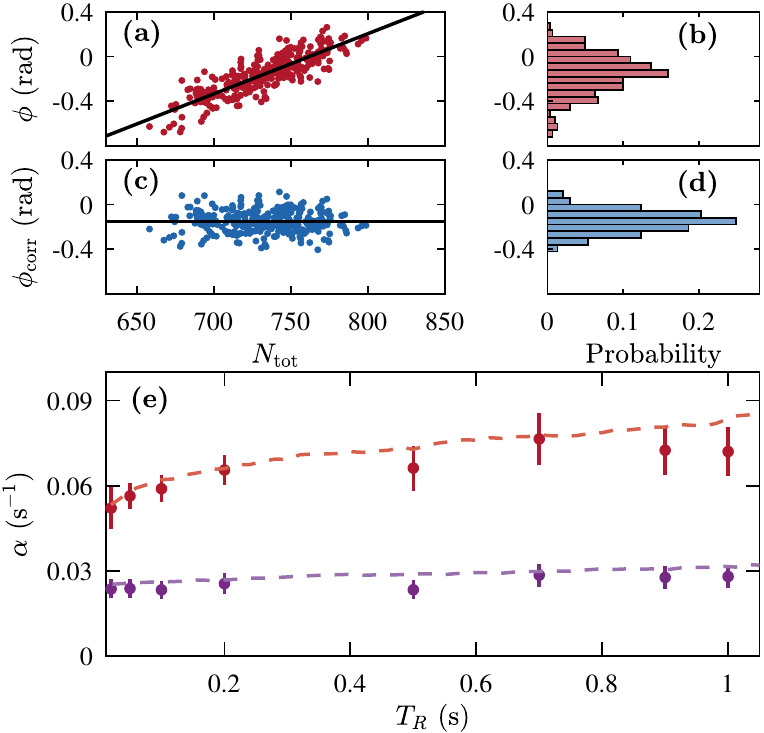}
\caption{\label{fig:chitilde}
Inteferometer data (a,c) and histogram (b,d) at fixed time $T_{R} =\unit{100}{ms}$ and fixed $\phi_{R}$, with the raw data (red) and data after the correction of collisional phase shift (blue). Black curves are a linear fit of the data. The raw $\phi$ shows a dependence on $N$, while after correction the slope is zero. (d) shows a reduced phase uncertainty compared to (b) due to the correction. (e) shows the collisional phase shift coefficient $\alpha$ at different times in a tight trap $\omega_{(r,z)}=2\pi\times(714,114) \unit{}{Hz}$ (red) and a shallow trap $\omega_{(r,z)}=2\pi\times(301,113) \unit{}{Hz}$ (purple). The dashed curves with corresponding colors show the results of the quantum trajectory simulation.}
\end{figure}

Both terms in the Hamiltonian will lead to a spread of the phase. The nonlinear term $\hbar\chi\Sz^2$ leads to one-axis twisting \cite{pezze2018}, sometimes also referred to as phase diffusion. However, in our case, low atom densities and nearly identical scattering lengths render its contribution very small with $\chi/2\pi\approx\unit{4.8\times10^{-4}}{Hz}$. The first term, which dominates in the Hamiltonian ($\tilde{\chi}/2\pi\approx\unit{8.6\times10^{-3}}{Hz}$), will introduce a phase evolution depending on the total atom number $N$. This so-called collisional clock shift has been studied both with BECs~\cite{hall1998, harber2002} and thermal clouds~\cite{clockshift1993, sortais2000, Rosenbusch2012}. As a result, for a fixed $T_R$, the measured phase $\phi$ shows a linear dependence on the total atom number $N$ (see Fig.~\ref{fig:chitilde}(a)), which can be extracted by fitting the data with the function
\begin{equation}
\label{eq:CS}
f(N)= \alpha(T_R)T_{R}N+\beta ,
\end{equation}
where $\alpha(T_R)$ and $\beta$ are free parameters. The value we obtain for $\alpha(T_R)$ can be used to quantify the strength of the collisional phase shifts. In the ideal case with $\chi=0$ and without atom loss, $\alpha(T_R) T_R N$ is proportional to $\int_{0}^{T_R}\tilde{\chi}(t)N(t)dt$. 

As is common also in atomic clock experiments, one can reduce the dephasing caused by atom number fluctuations by post-processing the data using the dependence of $\phi$ on $N$. The corrected phase $\phi_{\rm corr}=\phi-\alpha(T_R)T_{R}N$, as plotted in Fig.~\ref{fig:chitilde}(c), shows no dependence on $N$. This clock-shift correction allows to reduce the coherent effect and leads to a significant reduction of the observed phase noise, as shown in Fig.~\ref{fig:main-result}(c, d) and Fig.~\ref{fig:chitilde}(c, d). However, even after the correction, there is still significant residual phase noise, which can be attributed to the random nature of atom loss. Atom loss due to inelastic collisions occurs stochastically, thus randomizing the atom number time-evolution. In general, this leads to phase fluctuations since experimental shots with different atom number time-evolution will dephase with respect to each other. Experimentally, only the final atom numbers $N_1$ and $N_2$ are accessible, therefore the correction method described above does not allow to subtract the random collisional phase shifts completely.

To build a theoretical model taking into account this interplay of collisional interactions and the stochastic nature of atom loss, we model the quantum dynamics of the collective spin with the master equation \cite{alice1998}:
\begin{equation}
\label{eq:master}
\frac{\mathrm{d}\hat{\rho}}{\mathrm{d}t}= -\frac{i}{\hbar}\left[ \hat{H},\hat{\rho} \right] + \sum_{k=1}^4 \hat{C}_k\hat{\rho} \hat{C}_k^{\dagger}-\frac{1}{2}\hat{\rho}\hat{C}_k^{\dagger}\hat{C}_k- \frac{1}{2}\hat{C}_k^{\dagger}\hat{C}_k\hat{\rho} ,
\end{equation}
where the Hamiltonian is defined in Eq.~(\ref{eq:ham}).

The dominant types of losses are one-body losses described by loss constant $\Kone$, and the intrinsic losses from inelastic two-body collisions, described by inter-species and intra-species two-body loss constants $\KtwoAB$ and $\KtwoBB$, respectively. Two-body loss in state $|1\rangle$ and three-body losses are negligible for our parameters. The quantum jump operators $\hat{C}_k$ express the four types of significant losses in our system: one-body losses $(\hat{C}_1,\,\hat{C}_2) =( \sqrt{\Kone } \hat{a}_1,\,\sqrt{\Kone } \hat{a}_2)$ and two-body losses $(\hat{C}_3,\,\hat{C}_4) =( \sqrt{\gamma_{12}}\hat{a}_1 \hat{a}_2,\,\sqrt{\gamma_{22}}\hat{a}_2^2)$, where $\hat{a}_i$ is the bosonic operator annihilating atoms in the $\ket{i}$ state. Parameters $\gamma_{ij}$ are the integrated two-body loss rates, i.e. $\gamma_{ij} = \frac{K_{ij}^{(2)}}{2} \int\, {\rm d}^3 r\, |\psi_i|^2|\psi_j|^2$, where $|\psi_j|$ is the wavefunction of state $|j\rangle$. 

\begin{figure}
\includegraphics[width=8.6cm]{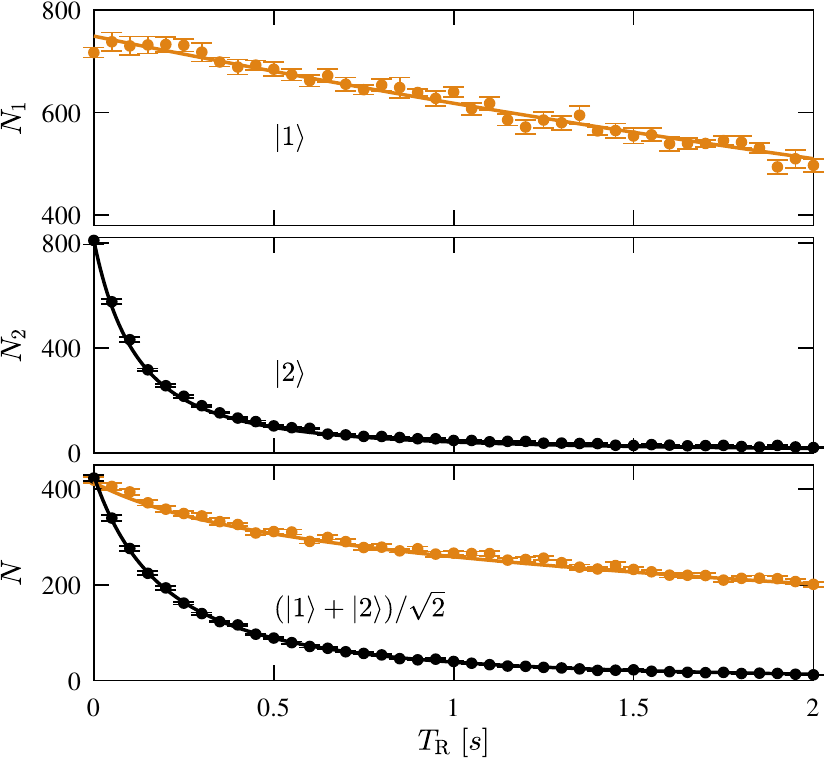}
\caption{\label{fig:AtomLoss}Time evolution of the mean atom number for three different initial states: all atoms in state $\ket{1}$ (top), all atoms in state $\ket{2}$ (middle) and atoms initially in the superposition $\bb{\ket{1}+\ket{2}}/\sqrt{2}$ (bottom). Orange (Black) points show the measured atom number in spin state $|1\rangle$ ($|2\rangle$). Error bars represent one standard error of the mean. Lines are from corresponding GPE simulations with loss rate constants $\Kone=0.17\unit{}{s^{-1}}$, $\KtwoBB=10.3\times10^{-14}\unit{}{cm^{3}s^{-1}}$ and $\KtwoAB=2.0\times10^{-14}\unit{}{cm^{3}s^{-1}}$.}
\end{figure}

In order to quantify the atom loss in our experiment, we perform three independent measurements. First, we prepare a BEC in state $\ket{1}$. With the loss of atoms happening solely due to one-body losses, we can extract $\Kone$ by a simple exponential fit, shown in Fig.~\ref{fig:AtomLoss}, top panel. In a second experiment, we measure the loss of atoms starting with a BEC in $|2\rangle$. In this case, there are one-body losses with already determined rate $\Kone$ and the intra-species two-body losses depending on $\KtwoBB$. We solve the one-component GPE including losses with varying $\KtwoBB$ to find the best fit to the experimental data, see Fig.~\ref{fig:AtomLoss}, middle. Eventually, we perform experiments with the BEC prepared in an equal superposition of $|1\rangle$ and $|2\rangle$ (Fig.~\ref{fig:AtomLoss}, bottom). In this case all loss channels are present, but only one unknown loss constant $\KtwoAB$ remains, which we determine by comparing the 2C-GPE simulation with the data. The loss rate constants determined in this way are $\KtwoBB=10.3(3)\times10^{-14}\unit{}{cm^{3}s^{-1}}$ and $\KtwoAB=2.0(1)\times10^{-14}\unit{}{cm^{3}s^{-1}}$. Our values differ from previously reported values in \cite{Egorov2013} but agree with the ones from \cite{van2002} and \cite{tojo2009}.

\begin{figure}[b]
\includegraphics[width=6.8cm]{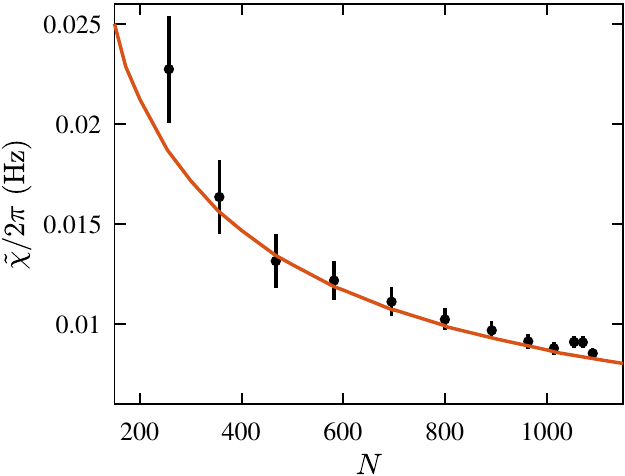}
\caption{\label{fig:scatteringlength} Comparison of the extracted initial $\tilde{\chi}$ between experiment (black dots) and 2C-GPE simulation (red line). For BECs with varying initial atom numbers we extract $\tilde{\chi}$ with a Ramsey sequence of $T_{R}=\unit{10}{ms}$.}
\end{figure}

Besides the loss rates, the phase noise evaluated from Eq.~(\ref{eq:master}) is also very sensitive to the precise values of the $s$-wave scattering lengths, especially the difference between $a_{11}$ and $a_{22}$, which affects the parameter $\tilde{\chi}$ crucially. To certify the values of scattering lengths used as input of the theoretical model, we run the Ramsey sequence with $T_{R}=\unit{10}{ms}$ with varying initial atom numbers and extract $\tilde{\chi}$. For this short interrogation time, atom losses are insignificant and $\alpha\approx\tilde{\chi}$. The 2C-GPE simulation with values of scattering lengths taken from \cite{Egorov2013} shows a good agreement with the experiment, see Fig.~\ref{fig:scatteringlength}. 

For long interrogation times the atomic densities decrease significantly due to atom loss. As a consequence, the parameters $\chi$, $\tilde{\chi}$ and $\gamma_{ij}$ also change with time. To take this effect into account, at each instant of time we compute these parameters numerically by solving {\it stationary} coupled Gross-Pitaevskii equations for the mean atom numbers $\overline{N}_1$ and $\overline{N}_2$, determined from Fig.~\ref{fig:AtomLoss}. Such an approach assumes that the atomic densities follow adiabatically the \cgpe~ground states and the superposition states with different atom loss trajectories evolve with the same parameters $\chi(t)$, $\tilde{\chi}(t)$, and $\gamma_{ij}(t)$. 

The phase noise simulation starts with an equal superposition of the two spin states. The initial number of atoms is drawn from a Gaussian distribution with mean and standard deviation as in the experiment. Its dynamics, given by Eq.~(\ref{eq:master}), is found with the help of the quantum trajectory method \cite{dalibard1992, carmichael1991}. This stochastic method results in a set of random final states $\left\{\ket{\tilde{\psi}( T_{R} ) }\right\}$, whose mixture is the solution of the master equation, i.e. $\sum \ket{\tilde{\psi}( T_{R} ) }\bra{\tilde{\psi}( T_{R} ) } \approx \hat{\rho}(T_{R})$. The numerically found stochastic wave-functions are treated as experimental realizations -- we analyze them in exactly the same manner as the real measurements to extract contrast, visibility, clock-shift correction and phase noise. Although the technical noise is very small, to ensure an appropriate comparison with the experiment, we add it to the simulation results. 

In Fig.~\ref{fig:main-result} we show the results of the quantum trajectory simulations. The simulation accurately reproduces the observed increase of phase noise, both before and after the clock-shift correction. We also show in Fig.~\ref{fig:chitilde}(e) the simulated and measured collisional phase shift as a function of time.

Our model does not include effects due to finite temperature, which can also lead to phase fluctuations \cite{Yvan2013,Yvan2018}. We neglect these effects because the absorption images do not show a thermal fraction or excitations in the gas for the $T_R$ investigated. Moreover, our zero-temperature model already accounts for the vast majority of the observed fluctuations, only for very long $T_R$ it slightly underestimates the phase noise. This shows that the phase coherence is limited mainly by the interplay of collisional interactions (elastic and inelastic) between atoms. 

To extend the coherence time, one can decrease the gas density, and thereby reduce the collisional rates. We confirm this by performing the phase noise measurement in a relaxed trap with trap frequencies $\omega_{\text{(r,z)}}=2\pi\times(301,113)\unit{}{Hz}$. As shown in Fig.~\ref{fig:main-result}(d) the phase noise is indeed strongly reduced. In order to reach the fundamental bound of precision in quantum metrology as theoretically described in~\cite{demkowicz2017adaptive, czajkowski2019many}, the effects of decoherence should be further mitigated. Experimentally, state-dependent potentials~\cite{li2009} could be applied to engineer interactions, so that the collisional phase shift is minimized or the quantum states are protected from the decoherence induced by two-body losses, as proposed in~\cite{Pawlowski2017}. 

In conclusion, we measure precisely the growth of the phase noise in a trapped two-component BEC and identify the main decoherence sources. We observe that the coherence is limited by random collisional phase shifts due to the stochastic nature of atom loss. The good agreement between our data and the GPE simulation also allows us to extract the two-body loss rate constants. Our experimental findings provide a good understanding of the temporal coherence of a two-component BEC. In contrast to most other systems where decoherence is due to interactions with the environment, in our experiment the atoms are well-isolated, and the observed decoherence effect is intrinsic to such a two-component BEC. Our findings are relevant for experiments on many-particle entanglement and quantum metrology~\cite{pezze2018} as well as trapped-atom clocks and interferometers where the atomic interactions play important roles~\cite{laudat2018}.

\begin{acknowledgments}
We acknowledge discussions with C. Salomon, A. Sinatra, and J. Reichel. We are grateful to Baptiste Allard for discussions and contributions to this work at an early stage. This work was supported by the Swiss National Science Foundation through NCCR QSIT and the (Polish) National Science Center Grant No. 2019/34/E/ST2/00289. Y. L. acknowledges support of a scholarship from the China Scholarships Council (No.201806210182). K. P. and P. T. acknowledge networking support by the COST Action CA16221. Center for Theoretical Physics, Polish Academy of Sciences is a member of the National Laboratory of Atomic, Molecular and Optical Physics (KL FAMO).
\end{acknowledgments}

\bibliographystyle{apsrev4-1}
\bibliography{CoherenceLimit}

\end{document}